\documentclass[aps,prl,twocolumn,superscriptaddress]{revtex4} 
\usepackage{amsmath}
\usepackage{graphicx} 
 \pdfoutput=1 

\begin{document}
\title{Superradiance mediated by Graphene Surface Plasmons}
\author{P. A. Huidobro}
\email{paloma.arroyo@uam.es}
\affiliation{Departamento de F\'isica Te\'orica de la Materia Condensada, Universidad Aut\'onoma de Madrid, E-28049 Madrid, Spain}
\author{A. Y. Nikitin}
\affiliation{Instituto de Ciencia de Materiales de Arag\'on and Departamento de F\'isica de la Materia Condensada, CSIC--Universidad de Zaragoza, E-50009 Zaragoza, Spain}
\affiliation{A. Ya. Usikov Institute for Radiophysics and Electronics, NAS of Ukraine, 12 Academician Proskura Street, 61085 Kharkov, Ukraine} 
\author{C. Gonz\'alez-Ballestero}
\affiliation{Departamento de F\'isica Te\'orica de la Materia Condensada, Universidad Aut\'onoma de Madrid, E-28049 Madrid, Spain}
\author{L. Mart\'in-Moreno}
\affiliation{Instituto de Ciencia de Materiales de Arag\'on and Departamento de F\'isica de la Materia Condensada, CSIC--Universidad de Zaragoza, E-50009 Zaragoza, Spain}
\author{F.J. Garc\'ia-Vidal}
\affiliation{Departamento de F\'isica Te\'orica de la Materia Condensada, Universidad Aut\'onoma de Madrid, E-28049 Madrid, Spain}

%\date{\today}

\begin{abstract}
We demonstrate that the interaction between two emitters can be controlled by means of the efficient excitation of surface plasmon modes in graphene. We consider graphene surface plasmons supported by either two-dimensional graphene sheets or one-dimensional graphene ribbons, showing in both cases that the coupling between the emitters can be strongly enhanced or suppressed.  The super- and subradiant regimes are investigated in the reflection and transmission configurations. Importantly, the length scale of the coupling between emitters, which in vacuum is fixed by the free space wavelength, is now determined by the wavelength of the graphene surface plasmons that can be extremely short and be tuned at will via a gate voltage. 
\end{abstract}

\maketitle

\section*{I. INTRODUCTION}
In the past few years, plasmonics has emerged as a way to control light at the subwavelength scale \cite{Barnes03,Maier,Bozhevolnyi}. Its potentiality is based on the properties of surface plasmons (SPs), which are surface electromagnetic (EM) waves coupled to charge carriers. These surface waves propagate along a metal-dielectric interface and are characterized by their subwavelength light confinement and long propagation lengths. Recently, it has been shown that graphene \cite{CastroNeto}, which has remarkable optical  \cite{Bonaccorso} and opto-electronic \cite{Liu} properties, also supports the propagation of SPs. These EM modes bounded to a graphene sheet have been studied theoretically \cite{Kenneth,Campagnoli,Vafek,Hanson,Jablan,Dubinov} and very recently observed in experiments \cite{Ju,Fei}. 

The properties of SPs in graphene have attracted great attention as they appear as an alternative for many of the functionalities provided by noble-metal SPs with the advantage of being tunable by means of a gate potential. For instance, by designing spatially inhomogeneous conductivity patters in a graphene sheet one can have a platform for transformation optics and THz metamaterials \cite{Vakil}. Additionally, strong light-matter interaction between SPs  and quantum emitters in graphene has been proposed based on the high decay rates of emitters close to graphene sheets \cite{Koppens11}. The field patterns excited by a nanoemitter in graphene were analyzed in Ref. \citenum{Nikitin11a}, demonstrating high field enhancements and long propagation lengths for the SPs. Furthermore, fluorescence quenching in graphene has been proposed as a probe of the evidence of plasmons \cite{Velizhanin,GomezSantos}.

The paper is organized as follows: in Sec. II we introduce the theoretical formalism and characterize the coupling between an emitter and the GSPs supported by a 2D graphene sheet. In Sec. III we study the interaction between two emitters mediated by the 2D GSPs. The presence of GSPs leads to interesting phenomena such as the super- and subradiant regimes, where the radiation properties of several emitters are enhanced or suppressed. We introduce the total normalized decay factor, $\gamma$, which characterizes the superradiant regime, and discuss its tunability. We analyze the evolution of $\gamma$ with the vertical distance between the emitters and the sheet in the transmission configuration, and with the in-plane distance between the emitters in the reflection configuration. In Sec. IV we consider the coupling of two emitters by GSP in one-dimensional (1D) graphene ribbons \cite{Nikitin11b}. We show that confining in 1D leads to longer interaction ranges between the emitters. Moreover, we study how the coupling between the emitters is affected by the presence of the ribbon's edges. Finally, our main results are summarized in Sec. V.

\section*{ II. COUPLING BETWEEN AN EMITTER AND GRAPHENE SURFACE PLASMONS}

\begin{figure}[ht]
  \centering
\includegraphics[width=1\linewidth]{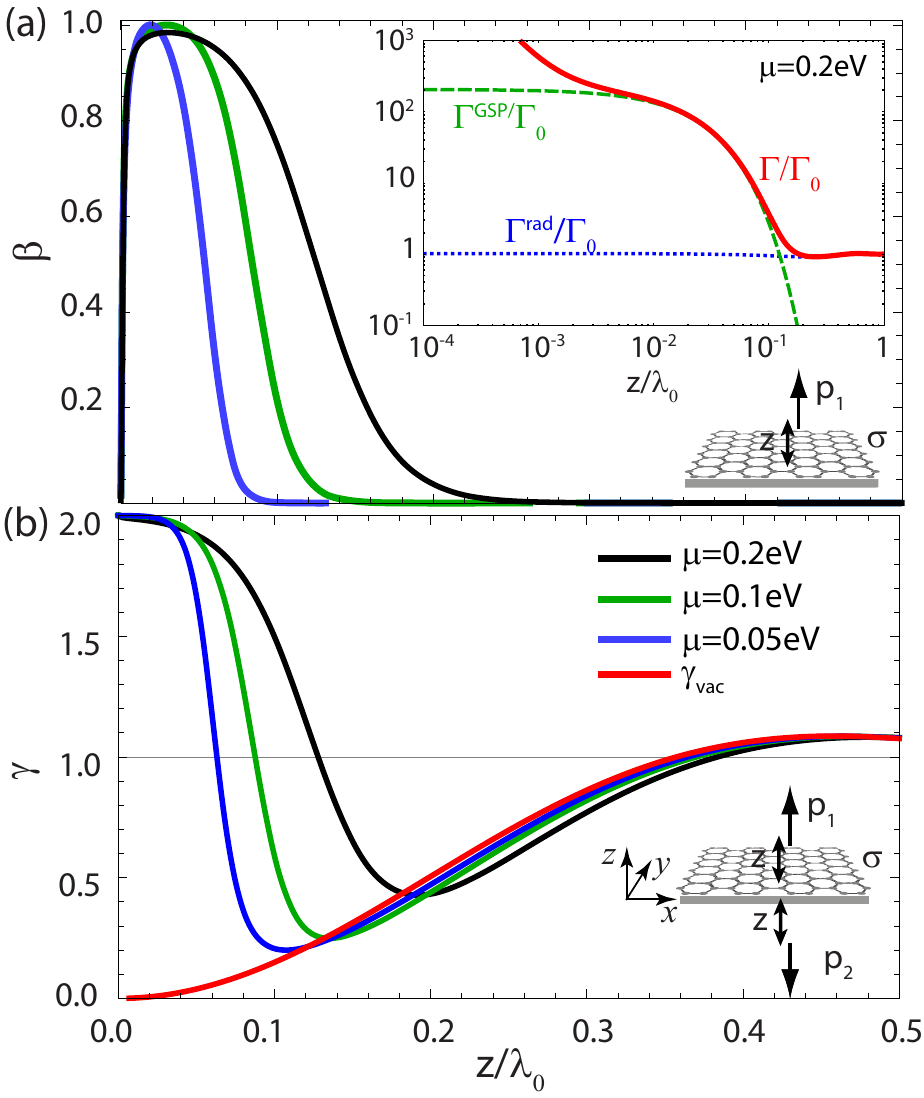}
  \caption{(a) $\beta$ factor for an emitter at 2.4THz as a function of the distance to the graphene sheet, $z$, normalized to the free-space wavelength, $\lambda_0$, for different values of the chemical potential ($\mu=0.2$, $0.1$ and $0.05$ eV). Inset panel: total decay rate ($\Gamma/\Gamma_0$) and decay rates through the plasmonic ($\Gamma^{GSP}/\Gamma_0$) and radiative($\Gamma^{rad}/\Gamma_0$) channels. (b) Super/subradiance between two emitters mediated by a graphene sheet when the two dipoles interact in transmission through it. $\gamma$ is plotted as a function of the vertical distance of the dipoles to the graphene sheet for the three values of the chemical potential $\mu$. The red line shows the vacuum interaction $\gamma_{vac}$. }
  \label{fig1}
\end{figure}

First, let us characterize the coupling to GSPs for an emitter at frequency $\omega$ decaying in the vicinity of a free-standing 2D graphene sheet. A sketch of the system under study can be seen in Fig. \ref{fig1} (a). The graphene sheet is placed in the $x-y$ plane and has a conductivity $\sigma(\omega)$, obtained in the random-phase approximation \cite{Wunsch,Hwang}. This quantity depends on the chemical potential of the graphene sheet, $\mu$, the temperature, which we consider to be $T=300$ K, and the carriers' scattering time $\tau$, for which we use a value taken from the theoretical predictions \cite{Hwang07b} such that $E_{\tau}=h/\tau=0.1$ meV. The emitter, modeled in the point dipole approximation, is placed at a distance $z$ from the sheet and has a dipole moment $\vec{p}$. 

When the emitter is close to the graphene sheet, it can decay through three different mechanisms: radiation to free-space, excitation of GSPs or coupling to absorption losses in the graphene sheet. The emitter's total decay rate is proportional to the imaginary part of the Green's tensor of the system, $\hat{G}(\vec{r},\vec{r},\omega)$, the dipole moment $\vec{p}$ and the free-space momentum $k_0=\omega/c$:
\begin{equation}
	\Gamma=\frac{2 k_0^2 |\vec{p}|^2}{\hbar \epsilon_0}\left\{ \vec{u}_{p}\mbox{Im}\left[\hat{G}(\vec{r},\vec{r},\omega)\right]\vec{u}_{p}\right\}
	\label{eqGamma}
\end{equation}
where $\vec{u}_{p}$ is the unitary vector in the direction of the dipole moment, For the sake of simplicity, we take a dipole moment perpendicular to the graphene sheet ($\vec{p}=p\vec{u}_z$), and then the relevant component of the tensor is $G^{zz}$ \cite{Nikitin11a}. This leads to the following expression for the Purcell factor, which is the total decay rate normalized to the free-space decay rate ($\Gamma_0=k_0^3 |\vec{p}|^2/(3\pi\hbar \epsilon_0)$):
\begin{equation}
	\frac{\Gamma}{\Gamma_0}=\frac{3}{2}\mbox{Re}\left[\int_0^{\infty}dq \frac{q^3}{q_z}\left(1-r_p(q) e^{2 i k_0 q_z z} \right)\right]
	\label{eqGamma11}
\end{equation} 
where we integrate over the normalized parallel wave vector $q=k_{\|}/k_0$, $q_z=\sqrt{1-q^2}$ is the momentum in the direction perpendicular to the sheet, with $\mbox{Im}(q_z)\geq 0$, and $r_p(q)=-\alpha q_z/(\alpha q_z+1)$ is the reflection coefficient of the graphene layer for p-polarization, with $\alpha=2\pi \sigma/c$ being the normalized conductivity. The pole of $r_p(q)$ gives the dispersion relation of the GSPs propagating in the graphene sheet, which appear when $\mbox{Im}(\sigma)>0$, i.e., below a critical frequency $\hbar\omega_0\approx 2\mu$. The contribution of GSPs to the total decay rate can be calculated from the pole in $r_p(q)$:
\begin{equation}
	\frac{\Gamma^{GSP}}{\Gamma_0}=\frac{3\pi}{2}\mbox{Re}\left[i \frac{q_p^2}{\alpha} e^{2 i k_0 q_z^p z}\right]
\end{equation} 
where $q_p=\sqrt{1-\alpha^{-2}}$ and $q_z^p=-\alpha^{-1}$ are the normalized momentum components of the GSP. 

The inset panel in Fig. \ref{fig1} (a) shows the Purcell factor (solid red line) at $2.4$ THz and for $\mu=0.2$ eV as a function of the emitter-graphene distance $z$ normalized to the free-space wavelength, $\lambda_0=124\mu$m. The physical parameters $\mu$ and $\lambda_0$ were chosen, as shown in Ref. \citenum{Nikitin11a}, to provide a good compromise in the trade-off between confinement and propagation length for the GSPs. Three different regions can be identified in the inset panel in Fig. \ref{fig1} (a) according to the decay mechanisms: (i) a radiative region at large distances ($z \geq\lambda_0/10$ for the chosen parameters), where the emitter is far enough from the graphene sheet and the total decay rate follows $\Gamma^{rad}/\Gamma_0$ (dotted blue line), which corresponds to the integration of the radiative modes in Eq. \ref{eqGamma11} ($0<q<1$); (ii) a region ($\lambda_0/10\geq z\geq\lambda_0/100$) where the dominant decay channel is the coupling to GSPs and the total decay equals $\Gamma^{GSP}/\Gamma_0$ (green dashed line); and (iii) a lossy region when the emitter is very close to the sheet ($z\leq\lambda_0/100$). Importantly, and as the figure shows, the decay rate of the emitter can be enhanced by several orders of magnitude. Here we are interested in the plasmonic region, where the GSP-contribution to the Purcell factor reaches values larger than $100$ for the parameters we have chosen. It is interesting to note that similar values of the Purcell factor can be obtained for very thin metal films \cite{Marocico} when the thickness is much smaller than the skin depth, which is challenging from the fabrication point of view, as opposed to graphene. For higher frequencies or smaller $\mu$, larger Purcell factors in the plasmonic region can be obtained in graphene: for instance, $\Gamma/\Gamma_0\approx 10^3$ at $\lambda_0=64\mu$m for the same chemical potential.

The parameter that accounts for the efficiency of the coupling to GSP, the $\beta$ factor, is defined as the ratio of the emitter's decay rate through GSP to its total decay rate, $\beta=\Gamma^{GSP}/\Gamma$. Fig. \ref{fig1} (a) studies the possibility of tuning $\beta$ with the chemical potential. Three values of $\mu$ are considered: $\mu=0.2$, $0.1$ and $0.05$ eV. For each value of $\mu$ there is a range of z's where $\beta$ is close to $1$, which corresponds to the region where the decay rate is dominated by the plasmonic channel (see inset panel). The region of high $\beta$ can be dynamically tuned with the chemical potential, which is in turn controlled by means of an electrostatic gating or a chemical doping. In particular, when the chemical potential is decreased to $0.1$ eV (green line) and $0.05$ eV (blue line), the GSP appears at larger q-vectors, the GSP is more confined to the graphene sheet, and the range of distances where $\beta$ is high is narrower. The capability of tuning plasmonic properties by means of a gate potential is the most important advantage of graphene compared to thin metal layers. 

\section*{III. SUPERRADIANCE IN TWO-DIMENSIONAL GRAPHENE SHEETS}
The efficient and tunable coupling of an emitter to the SP modes propagating in a graphene sheet can be used to modify the interaction between two emitters placed in the vicinity of the 2D sheet, similarly to SPs in metal surfaces \cite{MartinCano}. In order to study the GSP-mediated coupling between the two emitters, we introduce the normalized decay factor, $\gamma$, defined as the ratio of the total decay rate of a system where the two emitters interact through graphene to the decay rate of two uncoupled emitters in the presence of the graphene sheet. Thus, it accounts for the modification of the collective decay rate due to the presence of the second emitter and reads:
\begin{equation}
	\gamma=\frac{\Gamma_{11}+\Gamma_{12}+\Gamma_{21}+\Gamma_{22}}{\Gamma_{11}+\Gamma_{22}}
\end{equation}  
The decay rate $\Gamma_{ij}$ is the contribution to the decay rate of a dipole $\vec{p}_i$ placed at $\vec{r}_i$ due to the presence of a dipole $\vec{p}_j$ placed at $\vec{r}_j$ and can be written in terms of the Green's function that connects $\vec{r}_i$ to $\vec{r}_j$: $\Gamma_{ij}=2 k_0^2 |\vec{p}|^2/(\hbar \epsilon_0) \left\{ \vec{p}_{i}\mbox{Im}\left[\hat{G}(\vec{r}_i,\vec{r}_j,\omega)\right]\vec{p}_{j}\right\}$. Note that for $i=j$ we obtain the decay rate in Eq. \ref{eqGamma}, i.e., $\Gamma=\Gamma_{ii}$. The value of $\gamma$ characterizes two regimes: when $\gamma>1$ the interaction is enhanced due to the presence of graphene and the system is superradiant. Correspondingly, when $\gamma < 1$, there is an inhibition of the dipole-dipole interaction and the system is subradiant. Interestingly, a graphene sheet allows for two different configurations of the emitters: interaction in reflection (both emitters at the same side of the sheet) or in transmission (emitters placed at opposite sides).   

\begin{figure}[h]
  \centering
\includegraphics[width=1\linewidth]{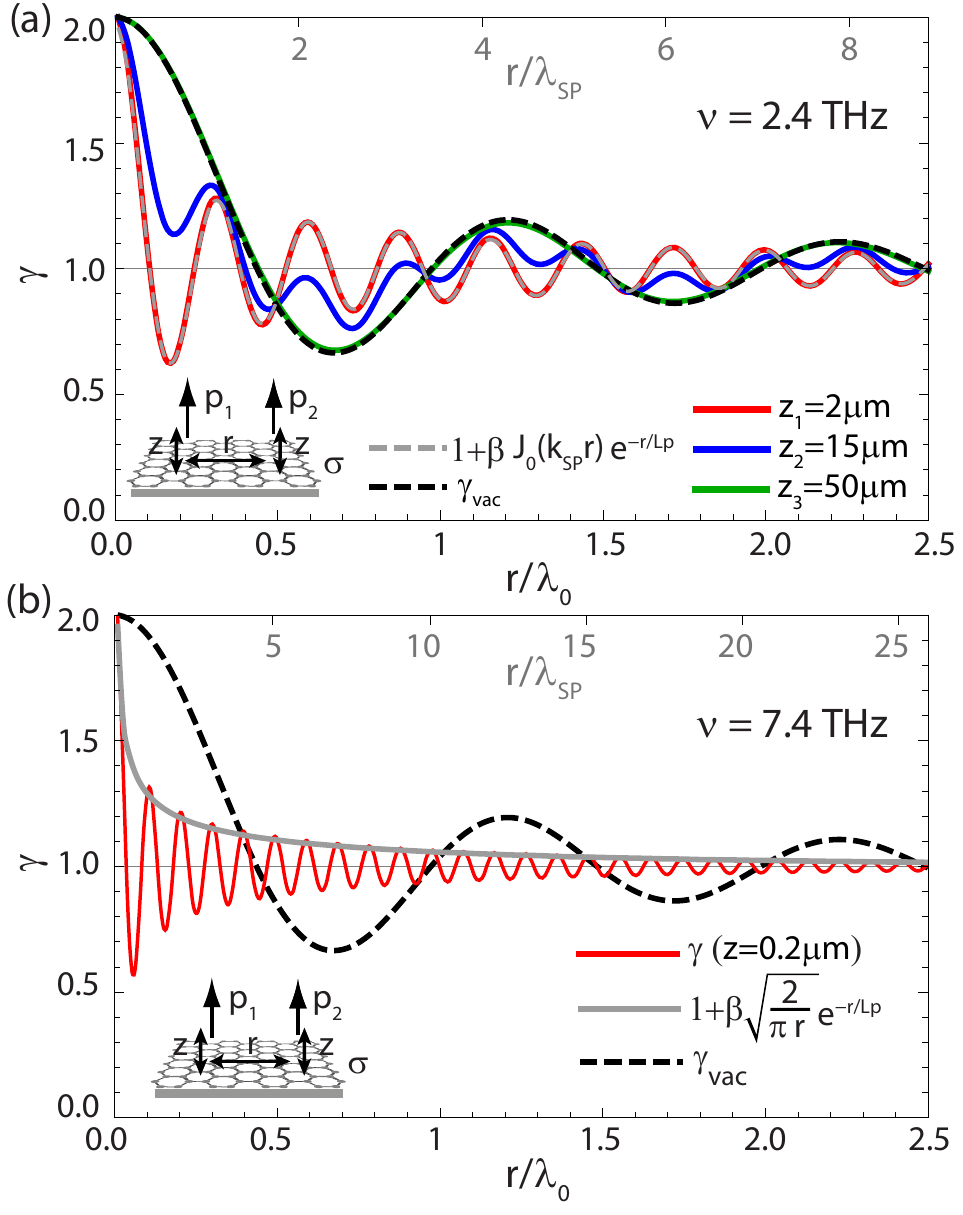}
  \caption{Tuning super- and subradiance: $\gamma$ factor as a function of the in-plane distance $r$ between two dipoles in the reflection configuration for two different frequencies. (a) At $\nu=2.4$ THz, when the dipoles are placed at $z_1=2\mu m$ ($ \beta=0.98$, $ \Gamma/\Gamma_0=105$), $z_2=15\mu m$ ($ \beta=0.55$, $ \Gamma/\Gamma_0=2$) and $z_3=50\mu m$ ($ \beta=0.01$, $ \Gamma/\Gamma_0=0.92$). The free-space wavelength is $\lambda_0=124\mu$m, the plasmon wavelength $\lambda_{p}=\lambda_0/3.5$ and the propagation length $L_p=14.9\lambda_{p}$.  The dashed black line corresponds to the free-space interaction and the dashed gray line to Eq. \ref{eqgamma} for high $\beta$. (b) At $\nu=7.4$THz, when the dipoles are at $z=0.2\mu m$, with $\beta=0.98$ and $\Gamma_{11}/\Gamma_0=2250$. The free-space wavelength is $\lambda_0=41\mu m$, the plasmon wavelength is $\lambda_{p}=\lambda_0/10$ and the propagation length is $L_p=20\lambda_{p}$. The dashed black line shows the free-space $\gamma$ and the gray line the decay of the interaction.}
  \label{fig2}
\end{figure}

Let us first consider two emitters interacting through the graphene sheet placed at opposite sides of the sheet, i.e., in the transmission configuration [see the sketch in Fig. \ref{fig1} (b)]. In order to study the behavior with the distance to the sheet, we locate the emitters at the same $z$ ($|z_1|=|z_2|$) and at $\vec{r}_{\|}=0$. We assume the dipole moments to be of the same module but anti-parallel, $\vec{p}_1=-\vec{p}_2=p\vec{u}_z$. The decay rate related to the interaction between the emitters, $\Gamma_{12}$, is needed to determine $\gamma$ and is related to the transmission coefficient:
\begin{equation}
	\frac{\Gamma_{12}^T}{\Gamma_0}=-\frac{3}{2}\mbox{Re}\left[\int_0^{\infty}dq \frac{q^3}{q_z} t(q) e^{2 i k_0 q_z z} \right]
	\label{eqGammaT}
\end{equation}
where the $-$ sign comes from the fact that the dipoles are anti-parallel and the transmission coefficient is $t(q)=1/(\alpha q_z+1)$. The $\gamma$ factor as a function of $z/\lambda_0$ is plotted in Fig. \ref{fig1} (b) for the same parameters used in panel (a). The red line shows $\gamma$ when the emitters are placed in free-space and interact only through radiative modes: $\gamma_{\mbox{vac}}=0$ at $z<<\lambda_0$ because the opposite phase of the dipole moments inhibits the radiation and when $z$ increases $\gamma_{\mbox{vac}}$ oscillates around $1$ with $\lambda_0$. When the graphene sheet is present, the interaction between the emitters is strongly modified at the subwavelength scale. In the limit of large $z$, in correspondence with the distances where $\beta\approx 0$ in panel (a), the emitters couple via radiative modes and $\gamma$ approaches $\gamma_{\mbox{vac}}$. On the other hand, in the range of $z$ where the plasmonic coupling between the emitters starts to dominate, $\beta\neq0$, $\gamma$ deviates from $\gamma_{\mbox{vac}}$, and, as the distance between the emitters and the sheet decreases, the system turns from subradiant to superradiant. For each value of $\mu$ ($0.2$, $0.1$ and $0.05$ eV), the value of $z/\lambda_0$ where $\beta$ starts to grow from 0 to 1, is the onset of the separation between $\gamma$ and $\gamma_{\mbox{vac}}$. Thus, the superradiant regime, controlled by high coupling to GSP, can be tuned by means of $\mu$. In the limit $z<<\lambda_0$, where $\beta=0$ again and losses dominate, the interaction reaches $\gamma=2$, in contrast to the free-space value in this limit, $\gamma_{\mbox{vac}}=0$. The reason for this is that, in this limit, the integrals in Eq. \ref{eqGamma11} and Eq. \ref{eqGammaT} are dominated by the contributions coming from large $q$, where $q_z=i|q_z|$ and $\Gamma_{11}$ is controlled by $-\mbox{Im}[r(q)]$ and $\Gamma_{12}^T$ by $-\mbox{Im}[t(q)]$. Since the imaginary part of both coefficients is the same, $\Gamma_{11}=\Gamma_{12}^T$ and thus $\gamma=2$. It is interesting to note that this sign change comes from the continuity conditions of the electromagnetic fields.

Let us now study how the interaction evolves with the in-plane distance between the emitters, $r=|\vec{r}_{1\|}-\vec{r}_{2\|}|$, where $\vec{r}_{\|}=(x,y)$. We take two dipoles interacting through the graphene sheet in the reflection configuration, as sketched in Fig. \ref{fig2} (a). In this case we place both of them at the same distance $z$ to the sheet, separated by an in-place distance $r$, and with dipole moments of the same magnitude, parallel and pointing in the vertical direction, $\vec{p}_1=\vec{p}_2=p\vec{u}_z$. In order to determine the $\gamma$ factor we need the interaction decay rate in reflection:
\begin{equation}
	\frac{\Gamma_{12}^R}{\Gamma_0}=\frac{3}{2}\mbox{Re}\left[\int_0^{\infty}dq \frac{q^3}{q_z} J_0(k_0 q r)\left( 1-r(q) e^{2 i k_0 q_z z} \right)\right]
\end{equation}
where the in-plane dependence is given by the zeroth-order Bessel function, $J_0$. The $\gamma$ factor as a function of $r/\lambda_0$ is plotted in Fig. \ref{fig2} (a) at 2.4THz and $\mu=0.2$ eV when the dipoles are at three different separations to the sheet, $z_1$, $z_2$ and $z_3$. First, for $z_1=2\mu$m, we know from Fig. \ref{fig1} (a) that the coupling to GSPs is very efficient, $\beta=0.98$, and $\Gamma/\Gamma_0=105$. In this situation, the interaction between the two emitters is mediated by GSP, and, consequently, the length scale of the interaction is controlled by $\lambda_{p}=\lambda_0/3.5$ as opposed to the free-space interaction, dominated by $\lambda_0$ (dashed line). When $\beta\approx1$, an analytical expression for $\gamma$ can be obtained in the pole approximation: 
\begin{equation}
	\gamma=1+\beta J_0(k_0 q_{p}r) e^{-r/L_p}
	\label{eqgamma}
\end{equation}
where $L_p$ is the propagation length of the GSP, given by $L_p=\lambda_0/[2\pi \mbox{Im}(q_{p})]$, and equal to $14.9\lambda_p$ in this case. As it can be seen in the plot, the analytical (gray dashed line) and exact (red line) calculations of $\gamma$ coincide. Since the propagation length is large enough, the decay length of the interaction is then given by the Bessel function, that decays as $\sqrt{2/(\pi r)}$. This is a dimensionality factor, coming from the fact that the 3D interaction in free-space is confined to the 2D graphene sheet. When the distance to the graphene sheet is increased, $\beta$ decreases and  $\gamma$ deviates from the analytical expression. For $z_2=15\mu$m the $\beta$ factor is $0.55$ and the shape of $\gamma$ reflects the fact that the emitter decays both to GSP and radiatively. Finally, when the distance to the sheet is large enough to have $\beta\approx0$, such as $z=50\mu$m, the vacuum interaction is recovered. Therefore, our results show that a larger interaction length scale and a modification of the super- and subradiant regimes can be achieved in a subwavelength scale for the appropriate choice of parameters.

The tunability of graphene enables us to reach a regime where the interaction between the two emitters can be controlled at very deep subwavelength scales. Although the tuning can also be done via $\mu$, here we show a situation where the tuning parameter is the frequency. When the two emitters interact in reflection [see Fig. \ref{fig2} (b)] at $7.4$ THz and $\mu=0.2$ eV, $\gamma$ (red line) is very different from the one corresponding to the free-space situation (black dashed line). Increasing the frequency while maintaining the chemical potential, results in a larger momentum for the GSP, $q_{p}$, which leads to a tighter confinement as well as a reduction in the propagation length. Thus, the interaction varies in a $\lambda_{p}=\lambda_0/10$ scale, as opposed to the free-space interaction, dominated by $\lambda_0$ [Fig. \ref{fig2} (b) shows both scales $r/\lambda_0$ and $r/\lambda_{p}$]. The decay of the $\gamma$ factor (gray line), is given by the square-root decay characteristic of 2D interactions for distances shorter than the propagation length, that in this case is $L_p=1.85\lambda_0$ ($19.2\lambda_{p}$). For larger distances, losses start to dominate and the interaction decays exponentially, according to Eq. \ref{eqgamma}.

\begin{figure}[ht]
\centering
\includegraphics[width=1\linewidth]{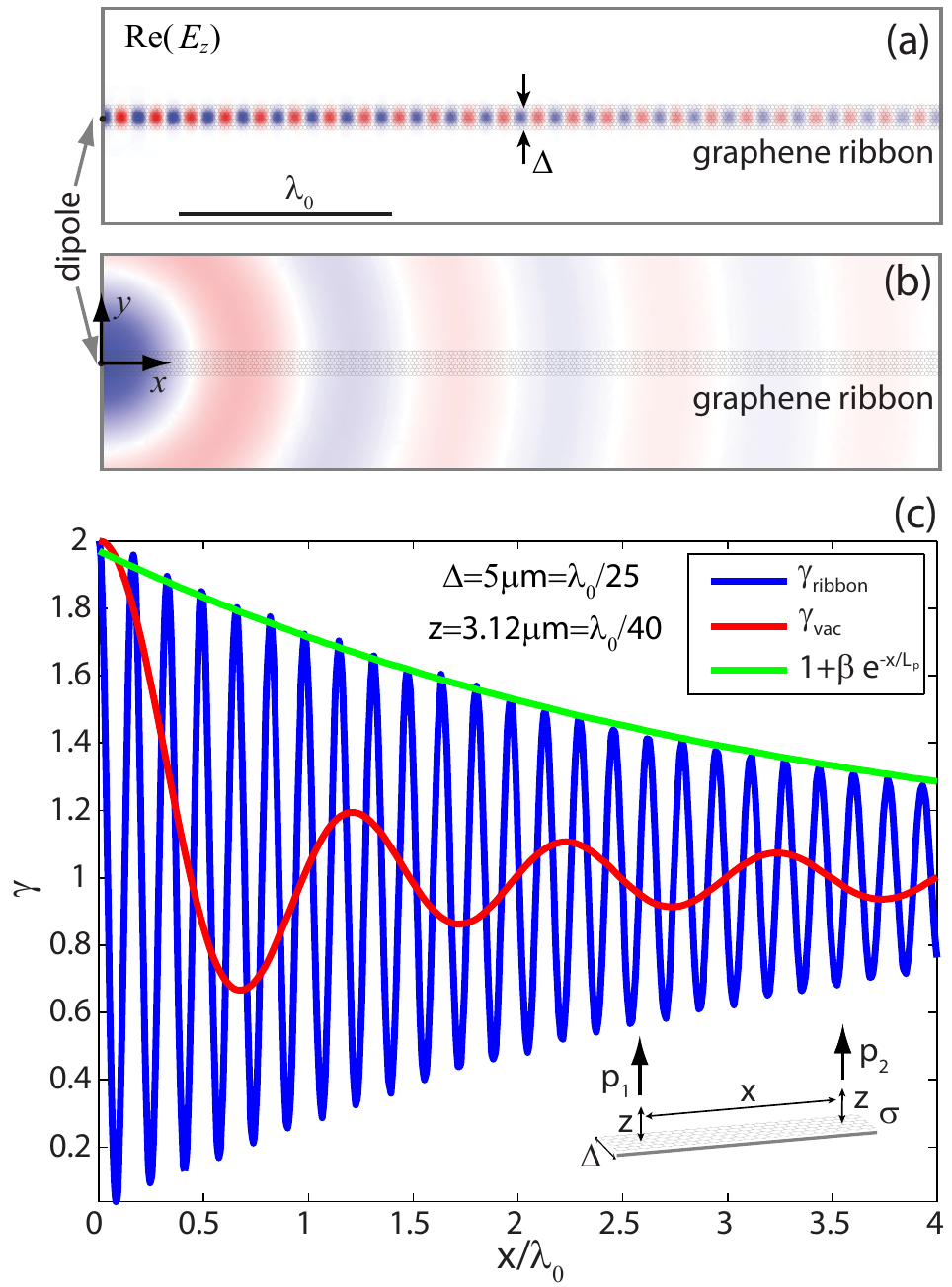}
\caption{Interaction mediated by graphene ribbons. (a,b) Electric field profile for a dipole decaying to the ribbon-GSP. The dipole is placed at $x=0$, $y=0$ and $z=\lambda_0/40$ in panel (a) and at $z=\lambda_0/10$ in panel (b) [the same would be obtained for $z>\lambda_0/10$]. (c) Super-radiance mediated by the ribbon-GSP mode shown in panel (a). The green line shows the exponential decay of the interaction.}
\label{fig3}
\end{figure}

\begin{figure*}
\centering
\includegraphics[width=1\linewidth]{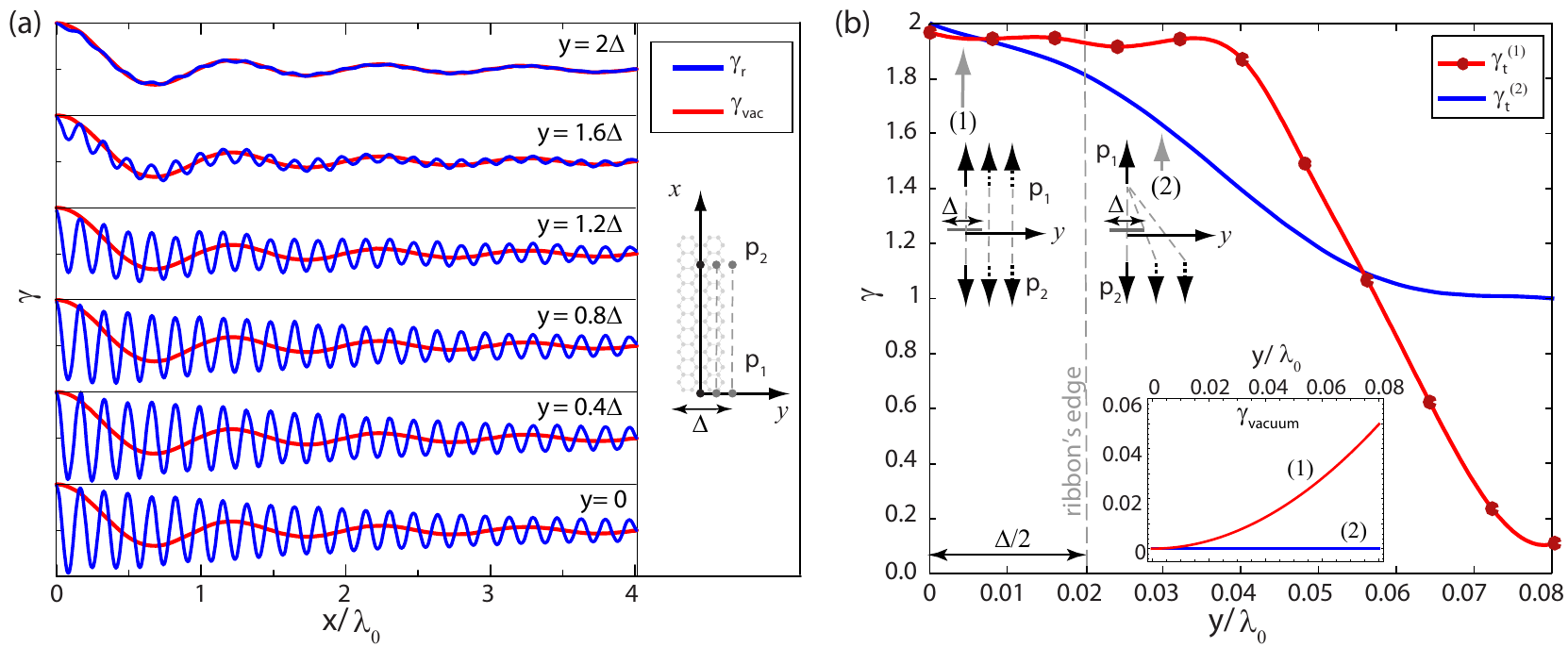}
\caption{Dependence of the dipole-dipole coupling in graphene ribbons on the lateral separation from the center of the ribbon, $y=0$. (a) $\gamma$ factor between two dipoles in the reflection configuration as a function of $x$ for several lateral separations from the center of the ribbon (blue lines). The scale in each sub-panel is between $0$ and $2$. The red line shows the corresponding interaction in vacuum. The inset panel shows a top view of the structure. (b) $\gamma$ factor as a function of the lateral separation $y$ in the transmission configuration for two cases. First (red line), $\gamma$ factor when the two dipoles are displaced simultaneously, as shown in sketch (1). The dots correspond to numerical simulations and the continuous line is a guide-to-the-eye. Second (blue line), $\gamma$ factor (from simulations) when the upper dipole is fixed at the center of the ribbon and the lower is displaced, as shown in sketch (2). The inset panel shows the corresponding interaction when the dipoles are placed in vacuum.}
\label{fig4}
\end{figure*}

\section*{IV. SUPERRADIANCE IN ONE-DIMENSIONAL GRAPHENE RIBBONS}
For completeness, we have also considered the possibility of confining GSPs in 1D graphene ribbons, which could provide a platform for long distance entanglement between two emitters, as proposed in Ref. \citenum{GonzalezTudela}. Compared to the GSPs propagating in a 2D graphene sheet at the same frequency, the ribbon-GSPs have a higher $q$ vector, thus it is more tightly confined to the graphene layer. On the other hand, while the coupling of two emitters mediated by 2D-GSPs is dominated by the dimensionality factor, $e^{-r/L_p}/\sqrt{r}$ (see Fig. 2), it is expected that in the case of 1D-GSPs this coupling will decay as $e^{-x/L_p}$, enabling long range interaction between the emitters provided $L_p$ is long enough. Let us consider a free-standing graphene ribbon of width $\Delta$ at $|y|<\Delta/2$, placed at $z=0$ with its axis along the $x$ direction (see Fig. \ref{fig3}). We take $\nu=2.4$ THz and the fundamental mode of a ribbon of width $\Delta=5\mu$m, that originates from the hybridization of two edge modes and has even parity of $E_z$ with respect to the ribbon axis \cite{Nikitin11b}. The field profile of the fundamental mode, obtained by means of the Finite Element Method (COMSOL software), is shown for two situations, where the emitter is placed at distances, $z=3.12\mu$m and $z=12.4\mu$m, as shown in Fig. \ref{fig3} (a,b). For a distance to the ribbon of $3.12\mu$m, i.e. $\lambda_0/40$, $\beta\approx1$, the GSP mode is excited with a very high efficiency and its field structure is clearly seen in panel (a). On the other hand, when the emitter is not sufficiently close to the ribbon, $\beta<<1$ and it couples mostly to radiation, as can be seen in panel (b) for an emitter at $z=\lambda_0/10$ where the field snapshot virtually coincides with an spherical wave. For the case with $\beta\approx1$, Fig. \ref{fig3} (c) shows $\gamma$ (blue line) for two emitters interacting in reflection through the ribbon-GSP. Similar to the 2D GSP, a subwavelength modification of the interaction can be achieved. However, since the ribbon-GSP is much more confined ($\lambda_{p}=\lambda_0/6.1$), this modification can be achieved in a shorter length scale. Moreover, propagation in 1D allows for a longer interaction range, with an exponential decay given by a propagation length $L_p=20\lambda_{p}$. Our results demonstrate that graphene ribbons could be used to control the length scale of the interaction between two emitters thanks to the efficient excitation of GSPs.

When analyzing graphene ribbons it is worth studying the transition across the the ribbon, i.e., how the coupling between two emitters is affected by the presence of edges (see Fig. \ref{fig4}). For this reason, we consider the evolution of the interaction in the reflection configuration, the same as in Fig. \ref{fig3} (c), and also plotting the evolution of $\gamma(x)$ as a function of the lateral distance from the center of the ribbon (see Fig. \ref{fig4} (a)). As sketched in the inset panel, we displace both emitters perpendicularly to the ribbon axis (along $y$), from the ribbon's center $y=0$ and passing through the ribbon's edge at $y=\Delta/2$, up to $y=2\Delta$.  As the figure shows, $\gamma (x)$ evolves from reflecting a high coupling to GSP at $y=0$, to following the free-space interaction at $y=2\Delta$, i.e., $\beta$ goes from $\approx 1$ to $\approx 0$ in a length scale of the order of $2\Delta$. Additionally, we also study the evolution of the interaction between two emitters with opposite dipole moments in the transmission configuration for two situations and plot $\gamma$ at $x=0$ as a function of $y$ in panel (b). First, we displace the emitters simultaneously such that both of them are at the same $y$ [see sketch (1)]. At $y=0$, the emitters couple to the ribbon-GSP and the system is in a superradiant state (red line), as opposed to the situation in free space, that is subradiant (inset panel, red line). When the emitters are displaced from the ribbon's center, but are still on top of the ribbon, i.e. $|y|<\Delta/2$, $\gamma$ is only slightly modified. Once the dipoles pass the ribbon's edge, the coupling to the 1D GSP is reduced, and therefore $\gamma$ decreases. Subradiance is quickly reached, approaching the free-space value, $\gamma=0$, which is achieved when the emitters are placed at a distance $2\Delta$ from the ribbon's center. In the second situation, one emitter is kept at the ribbon's center and the other is displaced perpendicularly to the ribbon's axis [see sketch (2)]. In this case, the system is always superradiant for the distances considered: $\gamma$ starts at $2$ and approaches $1$, while in free space $\gamma$ is of the order of $0.05$ at $y=\Delta$ (blue line in the inset panel). The reason for this lies on the fact that the emitter that is fixed always couples to the GSP. Remarkably, with only one emitter efficiently coupled to the ribbon-GSP, the interaction between both emitters is very different to the vacuum case. Our results for both configurations (reflection and transmission) demonstrate that the coupling between emitters mediated by 1D-GSP is very insensitive to the lateral displacement and that the effective lateral extension of these 1D-GSP is of the order of $\Delta/2$ measured from the edge's ribbon. 

\section*{V. CONCLUSION}
In conclusion, we have studied the tailoring of the interaction between two emitters mediated by surface plasmon modes in a graphene sheet. We have shown that within a certain range of distances to the graphene sheet, the decay rate of one emitter can be fully dominated by the graphene surface plasmons. Due to this efficient coupling,  the enhancement of the decay rate of the emitter, or Purcell factor, can be enhanced by several orders of magnitude. The interaction between two emitters mediated by the graphene plasmons in two-dimensional graphene sheets can thus be controlled at a subwavelength scale and can be tuned by means of external parameters. We have studied the appearance of the super- and subradiant regimes, both in the reflection and transmission configurations. Additionally, when the interaction is confined to one-dimension in graphene ribbons, a longer interaction range between the emitters and a very deep subwavelength control of the interaction can be achieved. Here, the lateral confinement leads to much higher enhancement factors, very deep subwavelength length scales for the coupling and longer interaction ranges. By considering the lateral displacement of the emitters from the ribbon's axis, we have also shown that the coupling to the graphene surface plasmons supported by the ribbon is very robust. Our results show that both graphene sheets and graphene ribbons can be used as efficient platforms to modify the interaction between two emitters when they are placed in their vicinity.  

%%%%%%%%%%%%%%%%%%%%%%%%%%%%%%%%%%%%%%%%%%%%%%%%%%%%%%%%%%%%%%%%%%%%%%%%%%
% Acknowledgements
%%%%%%%%%%%%%%%%%%%%%%%%%%%%%%%%%%%%%%%%%%%%%%%%%%%%%%%%%%%%%%%%%%%%%%%%%%
\section*{ACKNOWLEDGEMENT}
This work has been sponsored by the Spanish Ministry of Science and Innovation under Contract No. MAT2008-06609-C02 and Consolider Project Nanolight.es. P.A.H. acknowledges funding from the Spanish Ministry of Education through grant AP2008-00021 and A.Y.N. acknowledges the Juan de la Cierva Grant No. JCI-2008-3123.

%%%%%%%%%%%%%%%%%%%%%%%%%%%%%%%%%%%%%%%%%%%%%%%%%%%%%%%%%%%%%%%%%%%%%%%%%%
% References
%%%%%%%%%%%%%%%%%%%%%%%%%%%%%%%%%%%%%%%%%%%%%%%%%%%%%%%%%%%%%%%%%%%%%%%%%%


\begin{thebibliography}{99}

\bibitem{Barnes03} W. L. Barnes, A. Dereux, T. W. Ebbesen, Nature \textbf{424}, 824 (2003).

\bibitem{Maier} S. A. Maier, H. A. Atwater, J. Appl. Phys. \textbf{98}, 011101 (2003).

\bibitem{Bozhevolnyi} S. I. Bozhevolnyi, C. Genet, T. W. Ebbesen, Phys. Today (May) \textbf{44} (2008).

\bibitem{CastroNeto} A. H. Castro-Neto, F. Guinea, N. M. R. Peres, K. S. Novoselov, A. K. Geim, Rev. Mod. Phys. \textbf{81}, 109 (2009).

\bibitem{Bonaccorso} F Bonaccorso, Z. Sun, T. Hasan, A. C. Ferrari, Nat. Photonics \textbf{4}, 611 (2010).

\bibitem{Liu}  M. Liu, X. Yin, E. Ulin-Avila, B. Geng, T. Zentgraf, L. Ju, F. Wang, X. Zhang, Nature \textbf{474}, 64 (2011).

\bibitem{Kenneth} Kenneth W. -K. Shung, Phys. Rev. B \textbf{34}, 979 (1986).

\bibitem{Campagnoli} G. Campagnoli, E. Tosatti, in \textit{Progress on Electron Properties of Metals}, edited by R. Girlanda \textit{et al.} (Kluwer, Dordrecht, 1989), p. 337.

\bibitem{Vafek} O. Vafek, Phys. Rev. Lett. \textbf{97}, 266406 (2006).

\bibitem{Hanson} G. W. Hanson, J. Appl. Phys. \textbf{103}, 064302 (2008).

\bibitem{Jablan} M. Jablan, H. Buljan, M. Soljaci\'c, Phys. Rev. B \textbf{80}, 245435 (2009).

\bibitem{Dubinov} A. A. Dubinov, V. Y. Aleshkin, V. Mitin, T. Otsuji, V. Ryzhii, J. Phys. Condens. Matter \textbf{23}, 145302 (2011).

\bibitem{Ju} L. Ju, B. Geneg, J. Horng, C. Girit, M. Martin, Z. Hao, H. A. Bechtel, X. Liang, A. Zettl, Y. R. Shen, F. Wang, Nature Nanotech. \textbf{6}, 630 (2011). 

\bibitem{Fei} Z. Fei, G. O. Andreev, W. Bao, L. M. Zhang, A. S. McLeod, C. Wang, M. K. Stewart, Z. Zhao, G. Dom\'inguez, M. Thiemens, M. M. Fogler, M. J. Tauber, A. H. Castro-Neto, C. N. Lau, F. Keilmann, D. N. Basov, Nano Lett. \textbf{11}, 4701-4705 (2011).

\bibitem{Vakil}  A. Vakil, N. Engheta, Science \textbf{332}, 1291-1294 (2011).

\bibitem{Koppens11} F. H. L. Koppens, D. E. Chang, F. J. Garc\'ia de Abajo, Nano Lett. \textbf{11}, 11, 3370-3377, (2011).

\bibitem{Nikitin11a}A. Y.  Nikitin, F. Guinea, F. J. Garc\'ia-Vidal, L. Mart\'in-Moreno, Phys. Rev. B \textbf{84}, 195446 (2011).

\bibitem{Velizhanin} K. A. Velizhanin, A. Efimov, Phys. Rev. B \textbf{84}, 085401 (2011).

\bibitem{GomezSantos} G\'omez-Santos, G. and Stauber, T., Phys. Rev. B \textbf{84}, 165438 (2011).

\bibitem{MartinCano} D. Mart\'in-Cano, L. Mart\'in-Moreno, F. J. Garc\'ia-Vidal, E. Moreno, Nano Lett. \textbf{10}, 3129 (2010).  

\bibitem{Nikitin11b} A. Y. Nikitin, F. Guinea, F. J. Garc\'ia-Vidal, L. Mart\'in-Moreno, Phys. Rev. B \textbf{84}, 161497(R) (2011).

\bibitem{Wunsch} B. Wunsch, T. Stauber, F. Sols, F. Guinea, New J. Phys \textbf{8}, 318 (2006).

\bibitem{Hwang} E. H. Hwang, S. Das Sarma, Phys. Rev. B \textbf{75}, 205418 (2007).

\bibitem{Hwang07b} E. H. Hwang, S. Adam, S. Das Sarma, Phys. Rev. Lett. \textbf{98}, 98, 186806 (2007).

\bibitem{Marocico} C. A. Marocico, J. Knoester, Phys. Rev. A \textbf{84}, 053824 (2011).

\bibitem{Dzsotjan} D. Dzsotjan, J. Kastel, M. Fleischhauer, Phys. Rev. B \textbf{84}, 075419 (2011).

\bibitem{GonzalezTudela} A. Gonz\'alez-Tudela, D. Mart\'in-Cano, E. Moreno, L. Mart\'in-Moreno, C. Tejedor, F. J. Garc\'ia-Vidal,  Phys. Rev. Lett. \textbf{106}, 020501 (2011).  

\end{thebibliography}
\end{document}